\documentclass[12pt]{article}
\newcommand{\text}{\rm}
\begin{document}
\title{String Breaking and Z(2) Local Symmetry in the 3D 
Georgi-Glashow model}
\author{L.~E.~Oxman\\ \\Instituto de F\'{\i}sica, Universidade Federal Fluminense,\\
Campus da Praia Vermelha, Niter\'oi, 24210-340, RJ, Brazil.}
\date{\today}
\maketitle
\begin{abstract}
In this work, we consider the London limit of the $(2+1)$D Georgi-Glashow model 
with dynamical quarks. Following Polyakov's monopole plasma approach and using 
dual methods to treat the relevant matter sector, we derive in a clear and
straightforward 
manner the recently proposed gauging of the discrete Z(N) symmetry of the 
associated effective vortex theory. Our procedure applies to bosonic as well as to 
fermionic matter, enabling to derive a useful representation for the Wilson loop and 
discuss string breaking, due to the creation of dynamical quark-antiquark pairs. 
This phenomenon corresponds to a perimeter law, a behavior that has been already
observed in the lattice.
\end{abstract}

One of the most interesting problems we know at present in Physics corresponds to
look for possible phases and understanding confinement in strongly interacting systems 
like QCD. Then, the search for alternative methods enabling to describe at least the
relevant 
variables that determine the possible phases in the nonperturbative 
infrared regime becomes a central problem. 
One of these alternatives corresponds to duality; fundamental
progress in this direction has been developed many years ago by A. M. Polyakov and G. 't Hooft
\cite{polya,hooft1}.

In particular, the dual superconductivity scenario \cite{hooft1} can be realized in the
framework of the pure SU(N) Georgi-Glashow model in $(2+1)$ dimensions.
Classically, this model contains vortices with topological charge Z(N). At the
quantum level, the vortex sector can be represented by means of vortex operators
associated with instanton (monopole-like) singularities in euclidean spacetime,
where the vortices are created or destroyed. There are two relevant Green's functions: 
two-point and $N$-point correlators. They are incorporated  by means of 
the effective lagrangian for the vortex field \cite{hooft1},
\begin{equation}
\partial_\mu \bar{V}\partial^\mu V - \mu^2 \bar{V}V -\lambda (\bar{V}V)^2 -\beta 
(V^N +\bar{V}^N),
\label{vcond}
\end{equation}
which displays a global Z(N) symmetry (for a review, 
with further advances, see \cite{kovner} and references therein). In the Higgs
phase, when the vortex is an elementary 
excitation, the mass parameter $\mu^2$ is positive 
and there is no spontaneous symmetry breaking of the Z(N) group. Then, the vacuum
expectation 
value of the vortex field vanishes. However, when the vortices condense, the VEV of
the vortex field
is nonzero and Z(N) is spontaneously broken. This leads to domain walls between
different possible vacua which for planar systems are string-like. When
considering heavy probes, corresponding to a ``quark-antiquark'' pair, a string is
formed in 
between thus leading to an energy which is linearly increasing with the pair
separation, and an area law for the Wilson loop. This phase is then confining.

This nice mechanism continues to be an important topic of research; it is a defying
problem with 
regard to its possible implementation in QCD \cite{hooft1}, \cite{pwk}, 
while many open questions are still present in simplified models, which are also 
interesting in a condensed matter context \cite{kovner,igor-klein,mavro}.

Confining systems in $(2+1)$D that incorporate matter in the fundamental
representation are not so understood. 
However, the Wilson loop still provides information about the vacuum response to the
presence of a heavy quark-antiquark probe. 
In particular, a perimeter law for the Wilson loop would imply a constant potential, 
coming from the creation of dynamical light quark-antiquark pairs out of the vacuum. 
This phenomenon is called string breaking 
and has been observed in lattice QCD \cite{pw} (using the Polyakov loop) 
and lattice $(2+1)$D models containing bosonic \cite{b} or fermionic \cite{f}
dynamical matter. In particular, in ref. \cite{f}, string breaking has been 
detected by means of the Wilson loop criterium.

In ref. \cite{cesar}, by using the vortex 
condensation approach to the $(2+1)$D Georgi-Glashow model, the authors 
proposed the gauging of the Z(N) symmetry in eq. (\ref{vcond}), when bosonic matter 
in the fundamental representation is included. They also argued that 
the perimeter law is associated with a boundary effect coming from this gauging.

Besides the Georgi-Glashow model, confinement can be also understood in the case of
pure compact QED(3) \cite{polya}. In fact, both models are closely related as, in the 
London limit, the relevant low energy dynamics of the former is an abelian subgroup 
of SU(N), which is compact.

The aim of this letter is following the monopole plasma approach introduced by 
Polyakov to provide a simple derivation of the Z(2) gauging phenomenon
when dynamical fermionic quarks are present. Also,
upon obtaining a dual representation for the Wilson loop, we will discuss 
how the perimeter law is attained in the limit where the quarks become massless. \\

Then, our discussion will be focused on compact QED(3), eventually considered as
the London limit of the SU(2) Georgi-Glashow model, when the Higgs and $W^{\pm}$
become very massive. Let us take the following correlation function,
\begin{equation}
W(s)=\left\langle \exp i \int d^3x\, s_\mu\, f^\mu \right\rangle,
\label{ws}
\end{equation}
where $s_\mu$ is an external source to probe the system and $f_\mu =
\varepsilon_{\mu \nu \rho} 
\partial^\nu {\mathcal A}^\rho$ is the dual tensor for an abelian gauge field ${\cal
A}^\mu$.
The euclidean action associated with our model is,
\begin{equation}
\int d^ 3x\, \frac{1}{2} f_\mu f^\mu + ig \int d^ 3x\, J_\mu\, {\mathcal A}^\mu + K_F
\makebox[.5in]{,}K_F=\int d^3x\, \bar{\psi}(\partial \!\!\!/+m)\psi.
\label{system}
\end{equation}
In the context of the Georgi-Glashow model, $J^\mu$ and ${\mathcal A^\mu}$ refer to the
SU(2) 
current and gauge fields along the U(1) internal direction which is unbroken before 
compactification, while $\psi$ transforms in the fundamental representation of SU(2) 
and is formed by a pair of four component Dirac fields, describing a parity
symmetric system.

In particular, note that the correlator (\ref{ws}) will be useful to study the
Wilson loop,
$W=\left\langle \exp ig \oint dx_\mu\, {\mathcal A}^\mu \right\rangle$, integrated over
a boundary $\partial \Sigma$, as this loop can be rewritten, by using Stokes
theorem, in terms
of a 
($\Sigma$) surface integral of the dual tensor. That is, $W=W(g\, \delta_\Sigma\,
n)$, where 
$\delta_\Sigma$ is a Dirac $\delta$-distribution with support on $\Sigma$ and 
$n^\mu$ is the field of unit vectors normal to $\Sigma$, 
defined by the surface element $dS_\mu=n_\mu\, dS$. This is one of the steps we will
need
in order to compute the Wilson loop when compactifying the ${\mathcal A}$-sector of our
model, as this
process is accomplished by first performing the replacement,
\begin{equation}
f^\mu \rightarrow f^\mu + j^\mu
\makebox[.5in]{,}
j^\mu (x)= g_m\int_\gamma dy^\mu\, \delta^{(3)} (x-y),~~~ g\, g_m =2\pi,
\label{reple}
\end{equation}
where $\gamma$ is a Dirac string associated with instanton (monopole-like) 
singularities located at the string endpoints, and then
summing over 
all the instanton anti-instanton configurations (monopole plasma). In this manner, 
processes where ${\mathcal A}$-vortices are created or annhilated are incorporated. 
The vortex magnetic flux is given by the monopole magnetic charge $g_m$, 
which has been taken to saturate Dirac's quantization condition.

Then, we are left with the problem of implementing (\ref{reple}) in the second term
of the action (\ref{system}). For this aim, it is convenient
to use a dual transformation of the type introduced in refs. \cite{ibos,ch}, 
and extensively discussed in refs. \cite{ubos,univ,result3},
\begin{equation}
K_F+ig\int d^3x\, {\mathcal A}^\mu J_\mu \leftrightarrow
K_B+ig\int d^3x\, {\mathcal A}^\mu \varepsilon _{\mu \nu \rho }\partial ^\nu A^\rho, 
\label{bos}
\end{equation}
(see ref. \cite{univ})
where the U(1) fermion matter current is bosonized using a vector field $A_\mu$ 
whose gauge invariant action is given by,
\begin{equation}
e^{-K_B[A]}=\int {\mathcal D}b_\mu\, e^{-\Gamma [b]-\, i\int d^3x~b_\mu
\varepsilon^{\mu \nu \rho }\partial _\nu A_\rho }.  \label{sbosonic}
\end{equation}
The functional $\Gamma [b]=-\ln \det (\partial \!\!\!/+m+b \!\!\!/)$ is the effective 
fermionic action in the presence of an abelian gauge field $b_\mu$, along the
abovementioned 
unbroken U(1) direction. 
In this regard, note that if we exponentiate the second member in (\ref{bos}) 
and then use eq. (\ref{sbosonic}), we can path-integrate over $A$ to obtain a
$\delta$-functional 
leading to the constraint, $b_\mu=g {\mathcal A}_\mu + {\rm pure~gauge}$. 
Then, integrating over $b_\mu$ we are left with $\exp (-\Gamma [g {\mathcal A}])$, 
that is, the fermion path-integral using the first member in (\ref{bos}). 
Then, the mapping defined by eqs. (\ref{bos}) and (\ref{sbosonic}) 
is exact. The current bosonization rule $J_\mu \leftrightarrow \varepsilon _{\mu \nu
\rho} \partial ^\nu A^\rho$ is also universal, and it has been used to discuss the
universal transport properties in quantum Hall systems  (see ref. \cite{univ}).
This mapping generalizes the well known bosonization of $(1+1)$D fermionic 
systems, as in that case (\ref{bos}) and (\ref{sbosonic}) are valid with the 
replacement $\varepsilon _{\mu \nu \rho}\partial ^\nu A^\rho \rightarrow 
\varepsilon_{\mu \nu}\partial^\nu \phi$, that is, the usual bosonization rule for the 
fermion current, see also ref. \cite{vq}.

Now, in the $(2+1)$D case, because of the current bosonization rule, we can use an
integration 
by parts in the second member of (\ref{bos}) to rewrite the interaction in terms of
$f^\mu$, 
thus enabling a proper 
implementation of (\ref{reple}). Then, in compact QED(3) with dynamical fermions, 
the correlator (\ref{ws}) can be represented as,
\begin{equation}
W(s)=\frac{1}{{\mathcal N}}\int {\mathcal D}j\, {\mathcal D}A\, {\mathcal
D}{\mathcal A}\,
e^{-S+\, i \int d^3x\, s_\mu\, (f^\mu +j^\mu)},
\end{equation}
\begin{equation}
S= \int d^3x\, \frac{1}{2} (f +j)^2 + 
ig \int d^3x\, (f + j) A + K_B[A].
\label{act}
\end{equation}
The measure ${\mathcal D}j$ represents the path-integral over the monopole plasma,
whenever the integrand only depends on the monopole density. See ref. \cite{igor-klein}, 
for another possibility to implement the compactification in QED(3) models with 
dynamical matter; implications associated with both possibilities are interesting in
the context of applications to condensed matter systems \cite{igor-klein,mavro}.

In our case, we can follow the analysis for pure compact QED(3) given in ref.
\cite{antonov}, 
adapted for the computation of the correlation function $W(s)$.
Introducing a Lagrange multiplier $\lambda_\mu$ to linearize the Maxwell term in eq.
(\ref{act}),
and integrating over the field ${\mathcal A}^\mu$, we obtain the constraint, 
$g A_\mu -\lambda_\mu - s_\mu =\partial_\mu \phi$.  
Using this information in eq. (\ref{act}), we get,
\begin{equation}
W(s)=\frac{1}{{\mathcal N}}\int {\mathcal D}j\, {\mathcal D}A\, {\mathcal D}\phi\, 
e^{-\int d^3x\, \left( \frac{1}{2} (\partial\phi + s -g A )^2
+ i\, j^\mu \partial_\mu \phi\right) - K_B[A]},
\end{equation}
where we integrate over $\phi$, instead of $\lambda$. Then, the 
second term can be integrated by parts and written in terms of the monopole density
$\rho=\partial_\mu j^\mu$, so that the integration over $j$ is done over the 
monopole plasma, thus leading to a sine-Gordon term, $\cos g_m\, \phi$ \cite{polya}.
After the
field redefinition $g_m\, \phi \rightarrow \phi$, and using Dirac's quantization
condition 
$g\, g_m=2\pi$, we get,
\begin{equation}
W(s)=\frac{1}{{\mathcal N}}\int {\mathcal D}A\, {\mathcal D}\phi\, e^{ -\int d^3x\,
\left(\frac{1}{2g_m^2} 
(\partial\phi + g_m s - 2\pi A )^2
- \xi \mu^3 \cos \phi \right) - K_B[A]},\nonumber \\ 
\label{wdual}
\end{equation}
where $\xi$ is the monopole fugacity and $\mu$ is a constant with the dimension of
mass. 
Of course, if matter were not coupled, eq. (\ref{wdual}) would read,
\begin{equation}
W_0(s)=\frac{1}{{\mathcal N}}\int {\mathcal D}A\, {\mathcal D}\phi\, e^{ -\int
d^3x\, \left(\frac{1}{2g_m^2} 
(\partial\phi + g_m s )^2
- \xi \mu^3 \cos \phi \right)}.
\label{wpdual}
\end{equation}
It is worth underlining here that Polyakov's dual field $\phi$ can be identified 
with twice the phase of the vortex field $V$ (see \cite{kovner} and references
therein).
Then, the symmetry $\phi \rightarrow \phi +2\pi m$ in eq. (\ref{wpdual}) corresponds
to the 
Z(2) global symmetry of the $N=2$ effective vortex theory in eq. (\ref{vcond}).

When $s_\mu=g\, \delta_\Sigma\, n_\mu$, 
the correlator $W_0(s)$ gives the Wilson loop for pure compact QED(3), 
here represented by means 
of the celebrated dual action for the field $\phi$ (see ref. \cite{polya}).
According to Polyakov's analysis, the sine-Gordon term introduces a gap, 
leading to confinement of heavy probes. This comes about from a saddle point evaluation
of eq. (\ref{wpdual}),
\begin{equation}
\partial_\mu (\partial^\mu \phi + 2\pi \delta_\Sigma\, n^\mu) - \alpha^2 \sin \phi = 0
\makebox[.5in]{,}
\alpha^2=g_m^2 \xi \mu^3.
\end{equation}
Outside $\Sigma$, $\phi$ satisfies the sine-Gordon equation and is given 
by a domain wall containing a $2\pi$ discontinuity at $\Sigma$, coming from the 
source term. 
Therefore, when computing $(\partial\phi +g_m s)^ 2$ in eq. (\ref{wpdual}),
the associated delta singularity in $\partial \phi$ cancels against the $g_m
s=2\pi\,\delta_{\Sigma}\,n$ term. Away from $\Sigma$,
the field  is exponentially suppressed. Then, for a large surface, the domain wall
leads to 
an almost uniform action density localized on $\Sigma$ implying the area law for the
Wilson loop, that is, confinement.

Now, we see that in the case where the dynamical quarks are coupled, the theory is
strongly 
affected. Namely, the global Z(2) present in eq. (\ref{wpdual}) becomes
local. 
That is, the symmetry in eq. (\ref{wdual}) turns out to be,
$\phi \rightarrow \phi + 2\pi m(x)$, $A_\mu \rightarrow A_\mu + \partial_\mu m(x)$,
where $m(x)$ is an integer valued function.

According to the discussion in ref. \cite{kwp}, the main operational 
difference between the local and global versions of a discrete symmetry amounts to a
different prescription to carry out functional integrals. In the local case, the 
fields $\phi$ and $A_\mu$ take values in the quotient space. 
Then, with the equivalence $\phi\equiv \phi+2\pi$,
$\phi$ 
must be considered as an angle. 
As the dual action $K_B[A]$ is gauge invariant for any associated transformation 
$A\rightarrow A-(1/2\pi)\partial\phi$, we can decouple the $\phi$ field in the
correlator 
(\ref{wdual}). This also applies to the longitudinal component $\partial \psi$ of the 
external source $g_m s$ appearing in the correlation function (\ref{wdual}), 
\begin{equation}
g_m s_\mu= \partial_\mu \psi + R_\mu
\makebox[.5in]{,}
\psi(x)=-g_m\int d^3x'\, \frac{\partial'_\mu s^\mu(x')}{4\pi|x-x'|},
\label{dec}
\end{equation}
whenever $\psi$ corresponds to an ``angle'' field configuration. For  
the Wilson loop we have, $g_m \partial_\mu s^\mu = 2\pi\, n^\mu \partial_\mu
\delta_\Sigma$ 
and the values $\psi^{\pm}$, when we tend to a point on $\Sigma$ from above or
below, verify 
$\psi^+=\psi^- +2\pi$, which are identified. Then, we can absorbe the $\partial
\psi$ term in $W(s)$ by means of another transformation of the field $A$, and
 the correlator (\ref{wdual}) becomes,
\begin{equation}
W(s)=\frac{1}{{\mathcal N}}\int {\mathcal D}A\, e^{ -\int d^3x\, \frac{1}{2g_m^2}
(R-A )^2 - K_B[A]}.
\label{wla}
\end{equation}
As the system we are considering is parity symmetric (see below eq. (\ref{system})), 
$K_B[A]$ does not contain a Chern-Simons term; it only depends on the dual tensor
$F=\epsilon \partial A$. Then, transforming $A\to A+R$ and using eq. (\ref{dec}), the exponent 
in (\ref{wla}) will contain the term $\int d^3x\, \frac{1}{2g_m^2} A^2$, plus 
the bosonized action evaluated on $F+\epsilon \partial R=F+g_m\epsilon \partial s$ (cf. eq.
(\ref{dec})).

This leads to a qualitatively different situation when compared 
with the pure gauge result in eq. (\ref{wpdual}). There, the relevant source $s_\mu$ for 
the Wilson loop is concentrated on a surface, and the localization scale in the dual sine-Gordon
theory implies the area law. On the other hand, when dynamical quarks are present, the
local Z(2) symmetry changes the relevant source to be 
$\epsilon \partial s|^\mu=\epsilon^{\mu \nu \rho}n_\rho \partial_{\nu}\delta_\Sigma$, which 
is tangent to the boundary of $\Sigma$ and is concentrated there. 
For instance, in the case of a planar surface on the $x_3=0$ plane, the nonzero components are,
$+g \, \delta(x_3)\partial_2 \Theta_\Sigma$ and $-g \, \delta(x_3)\partial_1
\Theta_\Sigma$, for $\mu=1,2$, respectively, where $\Theta_\Sigma =1$, when the transverse 
coordinates $(x_1,x_2)$ are on $\Sigma$, and zero otherwise.

This change in the dimensionality of the source indicates that, depending on the localization 
properties in eq. (\ref{wla}), a perimeter law could be observed in the phase where 
dynamical quarks are coupled.

In order to go further, let us analyze the general structure of the fermionic effective 
action in eq. (\ref{sbosonic}). Again, because of parity symmetry, $\Gamma[b]$ depends on the dual
tensor $G=\epsilon \partial b$,
\begin{equation}
\Gamma [b] = \frac{1}{2}\int d^3x\, G O(-\partial^2) G +{\rm ~nonquadratic~part}.
\label{expa}
\end{equation}
where the nonlocal operator $O(-\partial^2)$ in the exact quadratic part comes 
from the exact vacuum polarization tensor. For large masses, it can be expanded according to, 
$O= (1/m)(c_1 + c_2\, (\frac{-\partial^2}{~m^2}) + c_3 (\frac{-\partial^2}{~m^2})^2
+\dots)$.
The $c_i$'s are adimensional constants with alternating signs. Because of charge conjugation 
symmetry, only terms with an even number of external legs contribute to $\Gamma[b]$. 
So that the nonquadratic part starts with four $G$ factors whose leading term is
$1/m^5$. In general, a finite order expansion of $O(-\partial^2)$ presents unphysical poles,
which are avoided by the implicit cut-off. This can be already seen at order $1/m^3$,
as $c_1>0$ and $c_2<0$. However, this can be improved by keeping the full momentum dependence of 
the polarization tensor (see ref. \cite{ubos}).

We will consider two important cases where additional information can be given: i) large $m$, where the nonquadratic part in $\Gamma$ is suppressed, ii) $m\to 0$; 
in this case, using renormalization group arguments, 
it has been argued that the infrared physics is associated with a nontrivial stable
fixed point, dominated by the quadratic term in eq. (\ref{expa}), where $O\sim \alpha\, (-\partial^2)^{-1/2}$
(see ref. \cite{igor-klein} and refs. therein). For a related discussion see also ref. \cite{FGM}.

Then, in these limiting situations, the (fixed point) quadratic fermionic effective action, 
together with eq. (\ref{sbosonic}), leads to a quadratic bosonized action,
\begin{equation}
K_B [A]= \frac{1}{2} 
\int d^3x\, F [-\partial^2 O(-\partial^2)]^{-1} F,
\label{quad}
\end{equation}
and path integrating in eq. (\ref{wla}), we arrive to an expression that can be used to study
the infrared Wilson loop behavior,
\begin{equation}
W(s)=  e^{ -\frac{1}{2}\int d^3x\, R\, [g_m^2+O]^{-1} R}=
e^{ -\frac{g_m^2}{2}\int d^3x\, \epsilon\partial s\, [(-\partial^2)(g_m^2+O)]^{-1}
\epsilon\partial s}.
\label{eh}
\end{equation}

In case i), using $O\sim (c_1/m) \to 0$ in eq. (\ref{eh}), we see that the exponent 
takes the form $(1/2g_m^2)R^2$.
This result is physically sensible: for any finite probe separation, if we start at a certain mass 
scale where string breaking occurs, as the fermion mass increases, a situation is expected 
where the dynamical quarks cannot produce a complete screening anymore.
This is manifested through the ``magnetic'' energy density form of the exponent, and 
the infinite localization scale in the field $R$,
\begin{equation}
R=\epsilon \partial [-\partial^2]^{-1} J
\makebox[.5in]{,}
J=g_m \epsilon \partial s,
\label{em}
\end{equation}
which prevents the onset of a perimeter law for any given finite loop.

When $m\to 0$, as discussed, we can still use eq. (\ref{quad}) to analyze infrared behavior.
For this aim, let us consider a Wilson loop, running along a large line of size $L$ contained on the 
$x_1$-axis, which then crosses a transverse distance of the order of $L$, goes along a line
parallel to $x_1$ and crosses back to close the loop.
The part on the $x_1$-axis gives a nonzero contribution $\epsilon \partial s|_1=g\delta(x_2)\delta(x_3)$
(see the discussion below eq. (\ref{wla})), whose 
associated Fourier transform is $g \sin(\frac{k_1 L}{2})/(\frac{k_1}{2})$. Then, in the
exponent of the second member in eq. (\ref{eh}), a typical relevant term giving the leading 
order behavior in $L$ is, 
\[
\frac{(g_m g)^2}{2} \int \frac{d^3k}{2\pi^3}\, \frac{1}{k^2(g_m^2+\alpha/k)} \left[ \frac{\sin(k_1 L/2)}{k_1/2}
\right]^2 \approx
\frac{Lg^2}{4\pi} \int dk_T\, (k_T+\alpha/g_m^2)^{-1},
\]
where $k_T=(k_2^2+k_3^2)^{1/2}$. We see that for small $k_T$
(large transverse distances) this result is finite, that is, no cut-off of the order of the loop
transverse dimension is needed. Therefore, when the complete loop is considered, 
the perimeter law is obtained. This corresponds to string breaking
due to the creation of light dynamical quark-antiquark pairs out of the vacuum.

Outside this limiting situations, the quadratic expansion of $\Gamma$ is not in
general a controlled procedure to study infrared behavior. Then, for finite masses, 
it would be interesting to include the effect of nonquadratic terms, in order to 
gain additional information about the scale for the onset of the Wilson loop 
perimeter law.\\

In this work, considering the London limit of the SU(2) Georgi-Glashow model, 
we have followed Polyakov's monopole plasma approach, together with dual methods 
for the fermionic matter, to derive in a straightforward manner 
a useful dual representation for the Wilson loop. In this way, a Z(2) local symmetry is directly 
evidenced. This gauging is realized in terms of the vector field
bosonizing the fermion current, along the U(1) subgroup of SU(2) which is unbroken before 
compactification.

Although we have considered fermionic quarks, represented by Dirac fields, 
our procedure can be also extended to scalar matter, by following the generalized dual
methods discussed in ref. \cite{du}.

The gauging of the Z(N) symmetry has been previously proposed in \cite{cesar}, 
using the vortex condensation approach to the SU(N) Georgi-Glashow model with
scalar matter in the fundamental representation. In that reference, 
a Wilson loop perimeter law finds an heuristic interpretation in terms of overlaps. 
Namely, when Z(N) is local, the Wilson loop operator changes the vacuum inside the
loop to a gauge equivalent state, while outside the loop it is left unchanged. 
Then, the overlap associated with the operator vacuum expectation value is expected 
to receive nontrivial contributions from a boundary effect.

Here, in the phase where dynamical quarks are coupled, we have seen that the theory displays a local 
Z(2) symmetry, and the relevant source is indeed concentrated on a boundary. 
In this phase, when the large mass limit is considered, the scale of localization 
becomes infinite and the perimeter law cannot be established for any given finite loop, as expected.
On the other hand, in the limit $m\to 0$, an infrared fixed point theory can be used to
study the Wilson loop behavior. In this case, a perimeter law is obtained, corresponding
to the creation of light dynamical quark-antiquark pairs.

Summarizing, string breaking is a widely observed phenomenon in lattice 
simulations of $(2+1)$D models. 
Therefore, it is desirable to obtain a general theoretical understanding of the
subject. Here, we have seen that the consideration of dual methods to treat the dynamical 
matter constitutes an interesting possibility to deal with compactification and discuss string breaking 
phases coming from the quark sector.

\section{Acknowledgements}
The author would like to thank C. Schat for useful comments.
The Conselho Nacional de Desenvolvimento Cient\'{\i}fico e
Tecnol\'{o}gico (CNPq-Brazil) and the Funda{\c {c}}{\~{a}}o de Amparo
{\`{a}} Pesquisa do Estado do Rio de Janeiro (FAPERJ)
are acknowledged for the financial support.


\end{document}